\documentclass[manuscript]{emulateapj}

\newcommand{\blast}{BLAST}
\newcommand{\rms}{r.m.s.}
\newcommand{\stn}{S/N}
\newcommand{\fwhm}{FWHM}
\newcommand{\bgunits}{\ensuremath{\mathrm{nW\,m}^{-2}\,\mathrm{sr}^{-1}}}
\newcommand{\microjy}{\ensuremath{\mu \mathrm{Jy}}}
\newcommand{\msol} {\ensuremath{\mathrm{M}_{\odot}}}
\newcommand{\tlb}{\ensuremath{t_\mathrm{lb}}}
\newcommand{\bzk}{\ensuremath{BzK}}
\newcommand{\inprep}[1]{#1 et al.\ (in prep.)}

\def\lsim{\mathrel{\lower2.5pt\vbox{\lineskip=0pt\baselineskip=0pt
           \hbox{$<$}\hbox{$\sim$}}}}

\def\gsim{\mathrel{\lower2.5pt\vbox{\lineskip=0pt\baselineskip=0pt
           \hbox{$>$}\hbox{$\sim$}}}}

\slugcomment{To appear in the Astrophysical Journal}

\shorttitle{BLAST: Resolving the CIB}
\shortauthors{Marsden,~G.~et al.}

\begin{document}


\title{BLAST: Resolving the Cosmic Submillimeter Background}

\author{
Gaelen~Marsden,\altaffilmark{1,\dag}
Peter~A.~R.~Ade,\altaffilmark{2}
James~J.~Bock,\altaffilmark{3}
Edward~L.~Chapin,\altaffilmark{1}
Mark~J.~Devlin,\altaffilmark{4}
Simon~R.~Dicker,\altaffilmark{4}
Matthew~Griffin,\altaffilmark{2}
Joshua~O.~Gundersen,\altaffilmark{5}
Mark~Halpern,\altaffilmark{1}
Peter~C.~Hargrave,\altaffilmark{2}
David~H.~Hughes,\altaffilmark{6}
Jeff~Klein,\altaffilmark{4}
Philip~Mauskopf,\altaffilmark{2}
Benjamin~Magnelli,\altaffilmark{7}
Lorenzo~Moncelsi,\altaffilmark{2}
Calvin~B.~Netterfield,\altaffilmark{8,9}
Henry~Ngo,\altaffilmark{1}
Luca~Olmi,\altaffilmark{10,11}
Enzo~Pascale,\altaffilmark{2}
Guillaume~Patanchon,\altaffilmark{12}
Marie~Rex,\altaffilmark{4}
Douglas~Scott,\altaffilmark{1}
Christopher~Semisch,\altaffilmark{4}
Nicholas~Thomas,\altaffilmark{5}
Matthew~D.~P.~Truch,\altaffilmark{4}
Carole~Tucker,\altaffilmark{2}
Gregory~S.~Tucker,\altaffilmark{13}
Marco~P.~Viero,\altaffilmark{8}
Donald~V.~Wiebe\altaffilmark{1,9} 
}

\altaffiltext{1}{Department of Physics \& Astronomy, University of
  British Columbia, 6224 Agricultural Road, Vancouver, BC V6T~1Z1,
  Canada}

\altaffiltext{2}{School of Physics \& Astronomy, Cardiff University, 5
  The Parade, Cardiff, CF24 3AA, UK.}

\altaffiltext{3}{Jet Propulsion Laboratory, Pasadena, CA 91109-8099,
  USA.}

\altaffiltext{4}{Department of Physics \& Astronomy, University of
  Pennsylvania, 209 South 33rd Street, Philadelphia, PA, 19104, USA.}

\altaffiltext{5}{Department of Physics, University of Miami, 1320
  Campo Sano Drive, Coral Gables, FL 33146, USA.}

\altaffiltext{6}{Instituto Nacional de Astrof\'isica \'Optica y
  Electr\'onica (INAOE), Aptdo. Postal 51 y 72000 Puebla, Mexico.}

\altaffiltext{7}{Laboratoire AIM, CEA/DSM-CNRS-Universit{\'e} Paris
  Diderot, IRFU/Service d'Astrophysique, B{\^a}t. 709, CEA-Saclay,
  F-91191 gif-sur-Yvette C{\'e}dex, France}

\altaffiltext{8}{Department of Astronomy \& Astrophysics, University
  of Toronto, 50 St. George Street Toronto, ON M5S~3H4, Canada.}

\altaffiltext{9}{Department of Physics, University of Toronto, 60
  St. George Street, Toronto, ON M5S~1A7, Canada.}

\altaffiltext{10}{University of Puerto Rico, Rio Piedras Campus, Physics Dept., Box
23343, UPR station, Puerto Rico 00931.}

\altaffiltext{11}{INAF, Osservatorio Astrofisico di Arcetri, Largo
  E. Fermi 5, I-50125, Firenze, Italy}

\altaffiltext{12}{Universit{\'e} Paris Diderot, Laboratoire APC, 10,
  rue Alice Domon et L{\'e}onie Duquet 75205 Paris, France.}

\altaffiltext{13}{Department of Physics, Brown University, 182 Hope
  Street, Providence, RI 02912, USA.}

\altaffiltext{\dag}{\url{gmarsden@phas.ubc.ca}}

\begin{abstract}
  The Balloon-borne Large Aperture Submillimeter Telescope (BLAST) has
  made one square degree, deep, confusion limited maps at three
  different bands, centered on the Great Observatories Origins Deep
  Survey South field.  By calculating the covariance of these maps
  with catalogs of 24\,\micron\ sources from the Far-Infrared Deep
  Extragalactic Legacy Survey (FIDEL), we have determined that the
  total submillimeter intensities are $8.60 \pm 0.59$, $4.93 \pm
  0.34$, and $2.27 \pm 0.20$\,\bgunits\ at 250, 350, and 500\,\micron,
  respectively. These numbers are more precise than previous estimates
  of the cosmic infrared background (CIB) and are consistent with
  24\,\micron-selected galaxies generating the full intensity of the
  CIB\@. 
  We find that the fraction of the CIB that originates from sources at
  $z\geq 1.2$ increases with wavelength, with 60\% from high redshift
  sources at 500\,\micron. At all BLAST wavelengths, the relative
  intensity of high-$z$ sources is higher for 24\,\micron-faint
  sources than it is for 24\,\micron-bright sources.
  Galaxies identified as active galactic nuclei (AGN) by their
  Infrared Array Camera (IRAC) colors are 1.6--2.6 times brighter than
  the average population at 250--500\,\micron, consistent with what is
  found for X-ray--selected AGN\@.  \bzk-selected galaxies are found to
  be moderately brighter than typical 24\,\micron-selected galaxies in
  the BLAST bands.  These data provide high precision constraints for
  models of the evolution of the number density and intensity of star
  forming galaxies at high redshift.
\end{abstract}


\section{Introduction}     
\label{sec:intro}

A decade ago, a uniform cosmic infrared background radiation (CIB) was
discovered in Far Infrared Absolute Spectrophotometer (FIRAS) data
from the {\em Cosmic Background Explorer\/} ({\em COBE\/}) by
\citet{puget1996}, and later confirmed by \citet{fixsen1998}. The
background, which peaks in intensity at a wavelength of $\approx
200$\,\micron, is as bright as all the optical light in the Universe
\citep{hauser2001}. It is presumed to be thermal re-radiation of
optical and UV starlight absorbed by dust grains.  At around the same
time as the discovery of the CIB, observations with the Submillimetre
Common User Bolometer Array \citep[SCUBA,][]{holland1999} at the James
Clerk Maxwell Telescope revealed a population of dusty high redshift
starburst galaxies forming stars at a rate of
${\sim}$\,1000\,\msol\ per year
\citep{blain1998,hughes1998,barger1998}.
 
Over the ensuing decade, a substantial effort has been devoted to
understanding the implications of this background radiation, and the
task has not been easy. Only the few very brightest submillimeter
galaxies are visible with SCUBA in blank-fields above the confusion
limit, and these comprise a small fraction of the total CIB at
850\,\micron\ \citep[e.g][]{hughes1998,scott2002,borys2003,coppin2006}.
Full understanding relies on statistical measures of submillimeter
galaxies too weak to be detected individually. Attempts to probe the
contribution to the CIB by faint sources have been made by looking at
objects lensed by clusters
\citep{cowie2002,smail2002,knudsen2008}. \citet{knudsen2008} have
resolved 100\% of the 850\,\micron\ background by reaching unlensed
flux density limits of 0.1\,mJy, but the reliability of the lensing
inversion introduces uncertainty in the estimate, and direct detection
of these sources is desirable. Furthermore, the CIB signal peaks at
wavelengths where the atmosphere is opaque and has dropped by almost
two orders of magnitude before it reaches the reliable atmospheric
window SCUBA exploits.  The bulk of the CIB is not visible from the
ground.
   
Since both the numerical density and the emission spectra of galaxies
evolve with cosmic epoch, the CIB is a convolution of these evolution
functions for a wide range of galaxy types. Because of this, the shape
of the CIB contains a wealth of information about the history of star
formation.  However, working from measurements at only one wavelength,
the convolution can not be inverted to extract an evolution function.
One expects a different mixture of galaxy types, and a different
redshift distribution, to emerge from observations of the CIB at
different wavelengths. Measurements at several wavelengths are
therefore required to constrain models of these underlying
distributions.
  
\blast, the Balloon-borne Large Aperture Submillimeter Telescope, was
designed to carry out the program of characterizing the CIB over a
range of wavelengths near to its peak.  We report here on the success
of that program.  \blast\ has produced deep, confusion-limited maps at
three wavelengths (250, 350, and 500\,\micron), where the CIB produces
substantial flux density, and we have used a uniform and carefully
constructed catalog of 24\,\micron-selected galaxies
\citep{magnelli2009} whose intensities we can find statistically in
the \blast\ data. These data bridge the gap between similar analyses
at shorter wavelengths, using data from the {\em Spitzer Space
  Telescope\/} \citep[e.g.][]{dole2006}, and at longer wavelengths,
using data from SCUBA \citep[e.g.][]{wang2006,serjeant2008}.
  
\citet{devlin2009} established that the full intensity of the CIB is
resolved statistically into flux density produced by identifiable
24\,\micron-selected galaxies.  In this paper we examine that relation
in more detail by dividing the 24\,\micron\ catalogs by brightness and
color.  A companion paper by \citet{pascale2009} uses spectroscopic
and photometric redshifts to constrain the star formation rate
history.  \inprep{Chapin}\ will explore the implications of
these new \blast\ results on our understanding of models of galaxy
evolution.

We analyze these high-\stn\ BLAST maps of confusion using techniques
that differ from those appropriate for catalogs of isolated point
sources; we use the covariances between our maps and external
catalogs, a technique known as ``stacking'', to measure the
background. Similarly, \citet{devlin2009} and \citet{patanchon2009}
use a ``$P(D)$'' fluctuation analysis, rather than counts of
individual sources, to provide the most accurate measurements of the
underlying source counts, while \citet{viero2009} study correlations
in the background. The possibility of determining number counts,
spectral energy distributions, and clustering biases directly from
correlations in the maps without the requirement to first extract a
catalog of point sources was pointed out by \citet{knox2001}.

The layout of this paper is as follows: in \S\,\ref{sec:observations}
the \blast\ observations and external data are described. In
\S\,\ref{sec:methodology} we develop the stacking formulae for
measuring the contribution to the total surface brightness detected in
a map produced at positions from an external catalog. Finally, in
\S\,\ref{sec:dissecting} we use 24\,\micron\ and optical catalogs, and
subsets based on color cuts, to determine the total CIB measured by
BLAST, as well as the relative contributions of low- and
high-redshift sources, AGN, and \bzk\ galaxies. Additionally, we
include an analysis of deep extragalactic SCUBA
850\,\micron\ observations, similar to those used by \citet{wang2006}.

\section{Observations}
\label{sec:observations}

\subsection{\blast}

\blast\ is a 1.8\,m stratospheric balloon-borne telescope that
operates at an altitude of approximately 35\,km, above most of the
opaque atmospheric water vapor that renders observations at all but a
few narrow submillimeter bands from the ground
impossible. Observations are undertaken simultaneously with three
broad-band bolometric imaging arrays with central wavelengths 250,
350, and 500\,\micron; this camera is a prototype of the Spectral and
Photometric Imaging Receiver (SPIRE) for {\em Herschel\/}
\citep{griffin2007}.

We use data from the 11-day BLAST flight in 2006 from McMurdo Station,
Antarctica. The under-illuminated BLAST primary produced nearly
diffraction-limited beams with full-width at half-maxima (FWHM) of
36\arcsec, 42\arcsec, and 60\arcsec\ in each band, respectively. This
successful flight produced significantly deeper- and wider-area
extra-galactic maps than existing 350 and 450\,\micron\ ground-based
observations, and the first 250\,\micron\ maps, near the peak in the
CIB. Deep and wide blank-field surveys were conducted in the Great
Observatories Origins Deep Survey South (GOODS-S) field, which we
label BLAST GOODS-S Deep (BGS-Deep) and BLAST GOODS-S Wide (BGS-Wide),
respectively. These maps are centered on the {\em Chandra\/} Deep
Field South. A coverage map is shown in \citet{pascale2009}. A second
intermediate-depth field near to the South Ecliptic Pole, which will
be the subject of a separate study, was also observed. In addition,
several low-redshift clusters and high-redshift radio galaxies were
targeted to sample biased star-forming regions of the Universe.
Further details on the instrument may be found in \citet{pascale2008},
and the flight performance and calibration for the 2006 flight are
provided in \citet{truch2009}.

In this paper we focus on the deepest \blast\ maps of
BGS-Deep\footnote{The \blast\ maps used in this paper are available
  for download at \url{http://blastexperiment.info/}.}
which cover an area of approximately 0.8 square degrees
\citep{devlin2009}, and completely encompass the Extended {\em
  Chandra\/} Deep Field South (ECDF-S), which in turn encompasses the
smaller GOODS-S and Hubble Ultra Deep Field South fields. The maps in
all BLAST bands are confusion limited, such that instrumental noise
itself contributes only a fraction (${\sim}\,$50\%, see below)
to the \rms\ of the map, and therefore provide high \stn\ measurements
of the spatial variations in the intensity of the CIB.

All BLAST time-stream detector data are reduced using a common
pipeline to identify spikes, correct time-varying detector
responsivities, and deconvolve the lag induced by thermal time
constants for the bolometers. Maps are produced from a combination of
these cleaned data with the pointing solution using a
maximum-likelihood algorithm \citep{patanchon2008}. The absolute
calibration is based on regular observations of the evolved star
VY~CMa, which results in uncertainties of approximately 10\% that are
strongly correlated between bands \citep{truch2009}. The resulting
maps may be thought of as the optimal weighting of the data on all
spatial scales such that both point sources and diffuse structures are
accurately reproduced within the limitations of the data. However,
while all spatial frequencies (inverse spatial scales) from the mean
level up to the Nyquist frequency are estimated, the largest scales
(including the absolute brightness) are not well constrained due to
detector drift and other systematics. We have therefore suppressed all
scales larger than approximately 9\arcmin, 7\farcm5, and
8\arcmin\ at 250, 350, and 500\,\micron, respectively --- this
procedure explicitly sets the mean of each map to zero.\footnote{The
  filter is, in practice, anisotropic, with greater suppression of the
  more poorly constrained modes orthogonal to the scan direction.}

\begin{figure}[tp]
\centering
\includegraphics[width=\linewidth]{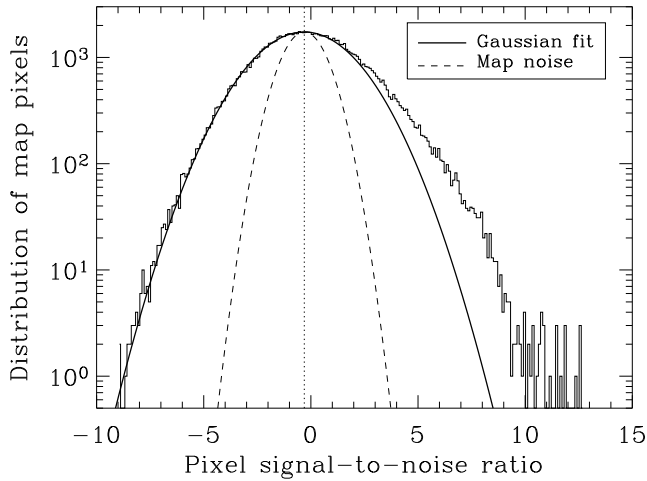}
\caption{The distribution of the pixel \stn\ ratio in BGS-Deep map at
  500\,\micron\ (histogram). The scale on the $y$-axis is the number
  of pixels per bin of width 0.1. The positive tail is due to bright
  sources in the map. Because the map has a mean of zero, the peak of
  the distribution is shifted to the left of zero, as indicated by the
  vertical dotted line. A Gaussian (thick solid curve) is fit to the
  negative side of the histogram. For reference, the measured
  instrument noise (a Gaussian, with width 1.0 in units of \stn) is
  over-plotted (thin dashed line). The map is clearly dominated by
  confusion noise. }
\label{fig:maphist}
\end{figure}

We show in Figure~\ref{fig:maphist} the noise properties of the
500\,\micron\ BGS-Deep map. 
We plot the distribution of the ratio of map pixel values to
instrumental noise, $\sigma_\mathrm{inst}$, estimated by the
map-maker, which propagates uncertainties in the time-domain detector
noise power spectra. The distribution is well-described by a Gaussian
with an excess at positive flux density, due to bright sources in the
map. The shape of the histogram is due to both instrumental noise and
confusion, $\sigma_\mathrm{conf}$. We estimate the confusion noise by
subtracting the instrumental noise from the map \rms\ in quadrature,
$\sigma_\mathrm{conf} = (\sigma_\mathrm{map}^2 -\,
\sigma_\mathrm{inst}^2)^{1/2}$. The map \rms\ is 16\,mJy and the
median instrument noise is 6.6\,mJy, thus we conclude that the noise
due to source confusion is $\sigma_\mathrm{conf}=15$\,mJy at
500\,\micron. Similar analyses give confusion noises of 17 and 21\,mJy
based on instrumental noises of 8.6 and 11\,mJy at 350 and
250\,\micron, respectively. The maps are clearly dominated by
confusion; this consists of a high \stn\ measurement of the shape
produced by galaxies much too close together to be resolved
individually.

\subsection{External Catalogs and Data}

In order to estimate the contribution of sources in the \blast\ maps
to the CIB, we stack the map at positions from catalogs measured at
other wavelengths. In the following subsections we describe these
external catalogs. The details of the stacking method are described in
\S\,\ref{sec:stacking}.

\subsubsection{FIDEL / SIMPLE}
\label{sec:fidel}

Recently it has been shown that deep Multiband Imaging Photometer for
{\em Spitzer\/} (MIPS) 24\,\micron\ catalogs contain significant
fractions of the sources that produce the CIB at 70 and
160\,\micron\ \citep{dole2006} and 850\,\micron\ \citep{wang2006},
wavelengths that bracket the \blast\ coverage. The deepest
24\,\micron\ catalog in BGS-Deep has been produced by
\citet{magnelli2009}, combining GOODS-S with Far-Infrared Deep
Extragalactic Legacy survey (FIDEL) data that cover a total of 0.21
square degrees. The FIDEL catalog was constructed using a
higher-resolution Infrared Array Camera (IRAC) catalog as a positional
prior. This deep catalog from SIMPLE \citep[{\em Spitzer\/} IRAC /
  MUSYC Public Legacy in ECDF-S,][]{gawiser2006} enabled the
deblending of sources with separations as small as 0.5\,\fwhm\ in the
MIPS map, resulting in an extremely faint flux density limit
$S_{24}\gsim 15$\,\microjy. The faintest source in the catalog has
$S_{24}=13\,$\microjy, but the detections at $S_{24} <
30$\,\microjy\ should be considered tentative. The effects of this are
discussed in \S\,\ref{sec:spurious}.

Furthermore, since the catalog is a 24\,\micron-detected subset of
SIMPLE, there also exist IRAC 3.6, 4.5, 5.8, and 8.0\,\micron\ flux
densities for each object. The depth, reliability, and wavelength
coverage of this catalog have been essential to this work.

\subsubsection{MUSIC}

Unlike the FIDEL catalog which has a fairly simple mid-IR selection
function, the MUltiwavelength Southern Infrared Catalog (MUSIC) uses a
near-IR selection with a much more heterogeneous data set combining
ACS and IRAC maps with ground-based $JHK$ Very Large Telescope (VLT)
data and optical spectroscopy \citep{grazian2006}. The primary use of
this catalog is to obtain spectroscopic and optical photometric
redshifts, although we also check the contribution to the CIB using
stacking.

\subsubsection{SCUBA} 

We extend the measurements made here with \blast\ to lower frequencies
using SCUBA data. The largest deep extragalactic SCUBA observations
are in the GOODS North (GOODS-N) field. Data from this field have
already been used to measure the 850\,\micron\ contribution to the CIB
\citep{wang2006, pope2007}, but, for consistency, we perform our own
analysis using techniques identical to what we have done with the
\blast\ data.  The map used here was produced by \citet{borys2003},
combining data from several different groups, and extended by
\citet{pope2005}, covering an area of approximately 0.06 square
degrees.\footnote{The SCUBA map of GOODS-N used in this paper is
available at
\url{http://www.noao.edu/staff/pope/DATA/SCUBA.html}.} The
noise in these maps is not uniform, varying from approximately 0.5 to
4\,mJy. The existing {\em Spitzer\/} catalogs in GOODS-N
\citep{papovich2004} are of similar depths to the FIDEL catalogs of
GOODS-S\@. The size of the SCUBA and 24\,\micron\ overlap region in
GOODS-N is 0.055 square degrees.

\section{Methodology}
\label{sec:methodology}

\subsection{Stacking} 
\label{sec:stacking}

Determining the mean flux density of a population of sources that are
individually too dim to be detected by examining their effect on a
confusion limited map is not new.  The main approach goes by the name
of ``stacking''.  Because this work makes such heavy reliance on the
technique, and because technical questions arise about the
generalization to very high source density or the wisdom of excluding
bright sources, etc., we review the basis of the technique here. We
find that many of these misconceptions are avoided when one realizes
that the technique is really one of taking the {\em covariance\/} of
the map with the catalog.

Imagine we have a map of the sky where $D^j$ is the flux density in
each pixel $j$ after convolution with the instrumental point-spread
function.  Suppose also that we have one or several independent
catalogs of sources made from other experiments; catalog $C_\alpha$
has $N_\alpha^j$ sources in pixel $j$, and we would like to know the
mean flux density, $S_\alpha$, of the sources in $C_\alpha$. We denote
the mean of $N_\alpha^j$ as $\mu_\alpha$, the average number of
sources per pixel in list $C_\alpha$. If all sources in a catalog
produce identical flux density, $S_\alpha$, then, along with whatever
else is in the sky, there will on average be a contribution of
$S_\alpha^j =S_\alpha N_\alpha^j$ to each pixel.

The \blast\ maps have zero mean.  We can write the flux density in the
map as
\begin{equation}
\label{eqn:map_flux}
D_j= S_1 (N_1^j -\mu_1) + S_2 (N_2^j -\mu_2)+ \ldots + w_j,
\end{equation}
where $w_j$ is the contribution of detector noise in pixel $j$, and,
strictly speaking, the $S_\alpha$ form the complete set of all objects
in the Universe. We have suppressed the mean by subtracting the
$S_\alpha\mu_\alpha$ for each catalog from every pixel. We
additionally require that $w_j$ has a mean of zero. We imagine
that the sources in the catalog are not correlated, such that
$N_\alpha^j$ is a random, Poisson-distributed number.  Furthermore,
assume that no two lists are correlated, so that
\begin{equation}
\label{eqn:no_corr}
\left<(N_\alpha^j-\mu_\alpha) (N_\beta^j-\mu_\beta) \right> =0, \,\,\, 
\forall \, \alpha\ne\beta.
\end{equation}

Our goal is to determine the mean \blast\ flux density per source in
list $C_\alpha$ from knowledge of the \blast\ maps, $D_j$, and the
locations, $N_\alpha^j$, of the sources in $C_\alpha$, but without any
other information, and in particular without requiring any knowledge of
other lists $C_\beta$ ($\alpha \ne \beta$).

$N_j$ is a function that has a shape on the sky, and measuring the
amplitude of this shape in the map determines $S_\alpha$, the mean
source brightness.  The covariance of $D_j$ with $N_j$ provides the
maximum likelihood estimate\footnote{This is the maximum likelihood
  estimate only if the map pixel noises $w_j$ are uniform. This is not
  the case for the \blast\ maps. We address this issue later on.} of
this amplitude:
\begin{eqnarray}
\label{eqn:covar}
\lefteqn{\mathrm{Cov}(D_j,N_\alpha^j)  =
\frac{1}{N_\mathrm{pix}}{\sum_j D_jN_\alpha^j} } \nonumber \\ 
 & = & \frac{S_\alpha}{N_\mathrm{pix}} \left( \sum_j (N_\alpha^j)^2 
  -\mu_\alpha\sum_jN_\alpha^j \right) \nonumber \\
 & = &  S_\alpha \sigma^2_{N_\alpha^j},
\end{eqnarray}
where $N_{\mathrm{pix}}$ is the total number of pixels in the map, and
we have dropped terms in $N_\alpha^jN_\beta^j$ and $N_\alpha^jw_j$,
which vanish in the sum. The first equality in Equation~\ref{eqn:covar}
holds because $D_j$ has a mean of zero, and the last equality follows
from the definition of variance. The variance of a Poisson
distribution is
\begin{equation}
\label{eqn:sigma}
\sigma^2_{N_\alpha^j}
= \mu_\alpha,
\end{equation}
so the net result is that the zero-lag cross-correlation (covariance)
of a catalog with the map divided by the mean number of sources per
pixel is an estimate of the average flux density per source,
\begin{equation}
\label{eqn:scon}
\hat S_\alpha =
{\mathrm{Cov}( D_j,N_\alpha^j)
  \over \mu_\alpha }.
\end{equation}

A final re-arrangement facilitates the use of Equation~\ref{eqn:scon}.
Notice that the sum in Equation~\ref{eqn:covar} runs over all pixels,
with the weight of each pixel proportional to the number of catalog
sources found within it, and that zero weight is given to pixels that
do not contain a catalog source ($N_j=0$).  This can be re-written as
a sum, instead, over all catalog entries with unity weight,
\begin{equation}
\label{eqn:catalog}
\hat S_\alpha
     =\frac{1}{N_\mathrm{pix}\,\mu_\alpha}{\sum_j D_jN_\alpha^j}
     = {1\over n_\alpha}{\sum_k D_k},
\end{equation}
where $k$ is the index of sources in catalog $C_\alpha$, $D_k$ is the
measured flux density in the map pixel that contains the
$k^\mathrm{th}$ catalog entry, and $n_\alpha$ is the total number of
catalog entries, $n_\alpha=N_\mathrm{pix}\,\mu_\alpha$. This
extraordinarily simple formula can be used to probe the properties of
sources much too crowded to be detected individually, and also those
with flux densities that are much fainter than the typical thresholds
of source catalogs derived only from the map itself. Since the map
pixel noises $w_j$ are not uniform across the map, we weight the mean
in Equation~\ref{eqn:catalog} by the inverse pixel variance to
maximize the \stn\ ratio of $\hat S_\alpha$. We explicitly subtract
the weighted means of the \blast\ maps.

Equation~\ref{eqn:catalog} provides a robust estimate of the mean
brightness per source even when there are other, possibly substantial,
contributors to the flux density present, $C_\beta$. This is provided
that $N_\alpha^j$ is Poisson distributed, and $N_\alpha^j$ is not
correlated with either the detector noise or sources in $C_\beta$.  In
other words, the effect of other sources on the estimator $\hat
S_\alpha$ is to provide an additional source of noise. This noise may
potentially be asymmetric, but it has a mean of zero, such that $\hat
S_\alpha$ is unbiased. 

Similarly, a catalog $C_\alpha$ can be subdivided into disjoint
subsets, and the mean brightness due to each subset can be measured
without bias. We use this fact to split up our catalogs based on IRAC
colors and brightness at 24\,\micron.
 
We are now in a position to address the proper handling of sources
that are bright enough to be easily recognized in the maps, for
example the sources in a \blast\ $5\,\sigma$ catalog.  We have shown
that $\hat S_\alpha$, our estimate of $S_\alpha$, is not affected by
either the presence or the removal of flux density from other source
lists $C_\beta$ that are uncorrelated with $C_\alpha$.  However, since
the sum of confusion noise and detector noise, $S_\alpha
N_\alpha^j+w_j$, will cause sources near the threshold to be
accidentally included or excluded from the \blast\ catalog, any list
made from the \blast\ maps themselves will be artificially correlated
with all the terms in Equation~\ref{eqn:map_flux}.  Furthermore, since
the \blast-generated bright source catalog depends on the sum of the
other terms in Equation~\ref{eqn:map_flux}, excision of the flux
density from such a catalog will artificially correlate the remaining
terms, such as $(N_\alpha^j-\mu_\alpha)$ and $w_j$.  This introduces a
bias in our estimator $\hat S_\alpha$ that is difficult to
quantify. In all the following work, stacking is performed on the full
\blast\ maps, including any bright sources they contain.

We reiterate that that this formulation for stacking is slightly
different than that used by other authors. In particular,
\citet{dole2006} performed aperture photometry on their stacked maps,
effectively subtracting a local background. The stacking analysis
performed by \citet{pascale2009} uses a similar technique to that of
\citet{dole2006}, and we have confirmed that their stacking
calculations give results consistent with those presented
here.\footnote{\citet{pascale2009} use a slightly different reduction
  and spatial selection of the \blast\ data than we use here, so we
  don't expect the results to be exactly the same.}

The relation between the average \blast\ flux density per source, $S_\nu$,
and the background specific intensity, $I_\nu$, is $S_\nu n/\Omega$,
where $\nu$ is the BLAST frequency and $n/\Omega$ is the number of
sources per solid angle. Note that the intensity of the background is
often defined through the product $\nu I_\nu$, which is then related
to $\nu S_\nu$. In what follows, we use $I=\nu I_\nu$. 

Near to its faintest limit, any catalog is only partially complete, so
we estimate $n/\Omega$ as $n_\mathrm{c}/(C\Omega)$ where
$n_\mathrm{c}$ is the number of catalog entries, $C$ is the
completeness of the catalog, assumed to be a function only of flux
density, and $\Omega$ is the solid angle of the catalog.  Completeness
of the FIDEL 24\,\micron\ catalog is measured by comparing the
differential flux density distribution of the catalog to published
completeness corrected counts.  Above 400\,\microjy\ we use
\citet[Table~2]{shupe2008} and between 35\,\microjy\ and
400\,\microjy\ we use \citet[Table~2]{papovich2004} after correcting
their flux densities for a 5.6\% difference in calibration that arises
from choices in {\em Spitzer\/} photometry methods.  The smooth
function $C= \left[ 1 +(A/S_{24})^\beta \right]^{-1}$, where
$A=16$\,\microjy\ and $\beta=1.8$, interpolates this correction. This
correction is $C=0.56$ at our lowest 24\,\micron\ flux density bin and
has a small effect on our final result (see \S\,\ref{sec:redshift}).

\subsection{Uncertainties}

\begin{figure*}[tp]
\centering
\includegraphics[width=\linewidth]{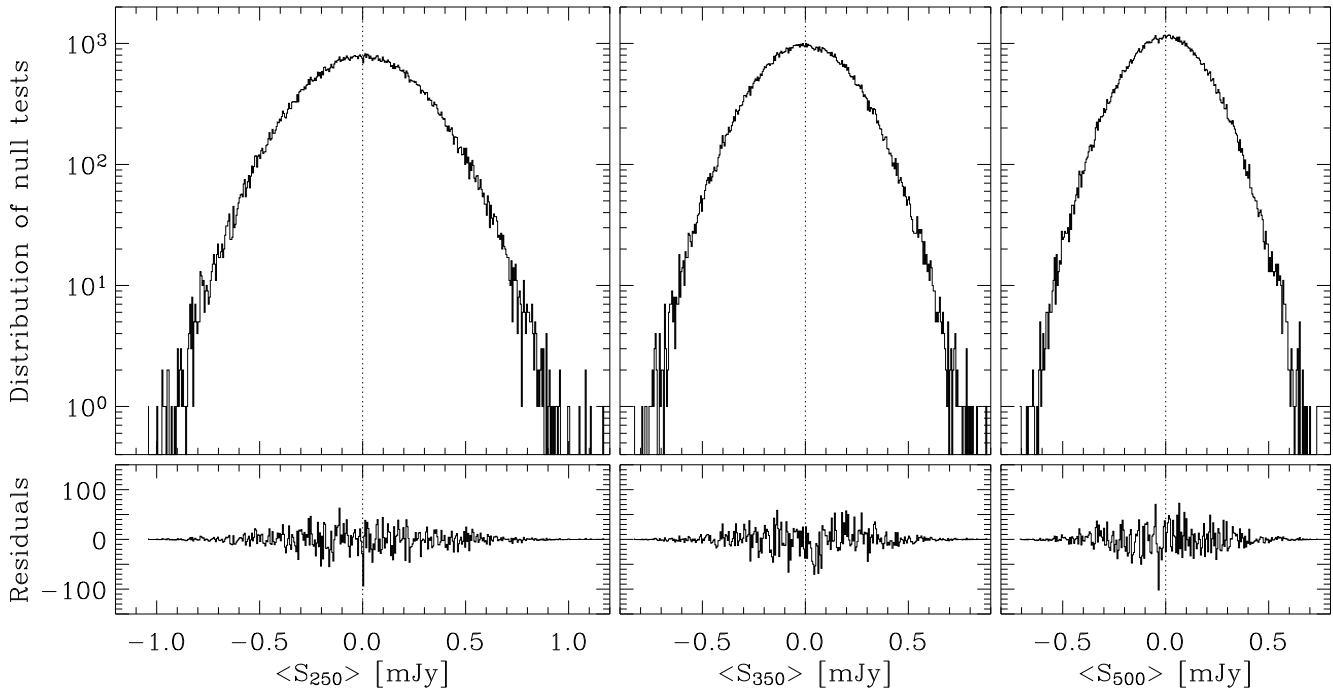}
\caption{Quantification of errors in FIDEL stacking measurements. We
  calculate the average BLAST flux density at $N$ random positions in
  the each map, where $N$ is equal to the number of sources in the
  FIDEL catalog. {\bf Upper row}: histograms of $10^5$ stacking
  measurements at random positions. The scale on the y-axis is the
  number of simulations per 5\,\microjy\ bin. {\bf Lower row}:
  difference between the histograms and Gaussians centered at 0.0 with
  width equal to the \rms\ of the distributions --- clearly the
  histograms are very well described by Gaussians. We use the \rms\ of
  these distributions as the errors in the stacked background
  values. We note that the errors between bands are strongly
  correlated, with Pearson correlation coefficients
  $\rho\,{\sim}\,0.7$ (see Table~\ref{tbl:firb}) in all cases.}
\label{fig:fidel_sim_err}
\end{figure*}

To estimate the uncertainty of Equation~\ref{eqn:catalog}
algebraically for a catalog $C_\alpha$, one would need to know the
scatter produced by its complement --- the catalog of all sources not
in $C_\alpha$ that contribute to the background (in addition to
sources of instrumental noise). In practice, the complement is not
known, so we establish the uncertainties and possible biases of our
measurements by generating random catalogs and stacking them on the
actual \blast\ maps under analysis.  We find, as expected, that the
uncertainties are Gaussian-distributed and scale as the map
\rms\ (including confusion noise) divided by the square root of the
number of catalog entries. See Figure~\ref{fig:fidel_sim_err}, and
notice the high precision of achieving a null, i.e.\ zero inferred
average flux density, from stacking on a catalog of random
positions. The uncertainties arise from sampling the maps at random
locations; since the confused flux density in the three separate BLAST
maps is correlated, the uncertainties for stacking are correlated
between BLAST bands. We measure the correlations from the simulations
and find $\rho_{12} = 0.70$, $\rho_{13} = 0.68$, and $\rho_{23}=
0.70$, where $\rho_{ab}$ is the Pearson correlation coefficient
between bands $a$ and $b$, and 1, 2, and 3 refer to 250, 350, and
500\,\micron, respectively.

\subsection{Limitations of the method}
\label{sec:limitations}

The stacked signals only return the mean source flux density if the
catalog is Poisson distributed, since we have used the fact that the
variance of the number of catalog entries in a region of a given size
equals the mean.  When this assumption is violated, the relation
between the covariance of a map with a catalog and the corresponding
mean flux density becomes complicated.  If the catalog $C_\alpha$ is
clustered at the scale of the \blast\ beams, it is very easy to
overestimate the mean flux density $S_\alpha$ and therefore
overestimate the contribution of a given catalog to the total sky
intensity.

This problem is strikingly apparent when we perform the covariance
with the MUSIC catalog, which has ${\sim}\,$18{,}000 sources in 0.035
square degrees (500{,}000 sources per square degree). We find that the
background intensity inferred by stacking the \blast\ maps on the
MUSIC catalog exceeds the FIRAS values by factors of 2.5--3.0. We
suggest that clustering in the catalog (whether due to selection
biases or real clustering on the sky) is the cause of this
overestimate.

The level of clustering that is measurable in a given area is strongly
dependent on the source density, since, for a given level of
clustering, the accuracy with which the mean can be measured goes as
the square root of the number of sources in that area. We show here
that the MUSIC catalog is strongly clustered at the scale of the
\blast\ beams and thus we should expect a biased measurement of the
background, but that the FIDEL catalog, with ${\sim}\,$9100 sources in
0.21 square degrees (43{,}000 per square degree), is well-behaved at
the appropriate scales.

\begin{figure}[tp]
\centering
\includegraphics[width=0.8\linewidth]{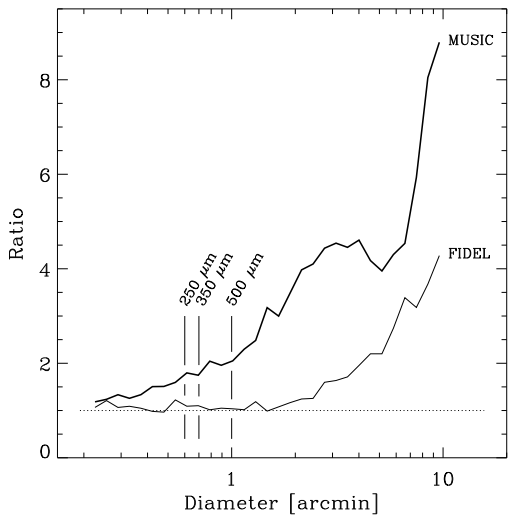}
\caption{Clustering in the MUSIC and FIDEL catalogs as a function of
  scale. The ratio of the variance of the number of sources to the
  mean number of sources within a circle is plotted against the
  diameter of the circle for both the MUSIC and FIDEL catalogs.  For a
  Poisson-distributed catalog, this ratio would be unity at all
  scales.  Each point in the graph is the variance (divided by the
  mean) of the number of entries that fall within 500 randomly-placed
  circles of a given size within the respective survey area. The
  broken vertical lines indicate the BLAST FHWMs. Notice that the
  FIDEL catalog retains a Poisson distribution to larger scales than
  the BLAST beams while the much deeper and heterogeneous MUSIC catalog
  shows excess variance arising from the clustering of galaxies at all
  angular scales.}
\label{fig:poisson_compare}
\end{figure}

We have measured how close to Poisson-distributed the MUSIC and FIDEL
catalogs are.  We place 500 circles of diameter $D$ at random within
the catalog area and count the number of catalog entries in each.
Figure~\ref{fig:poisson_compare} shows the ratio of the variance to
the mean as a function of circle diameter, for MUSIC and FIDEL.
If the catalogs were uniform random distributions, this ratio would be
unity.  The MUSIC catalog shows a substantial excess variance at all
angular scales above a few tens of arc seconds.  It is not a surprise
that galaxy locations are correlated and that MUSIC is deep enough to
measure that. We also note that MUSIC is an extremely heterogeneous
catalog, consisting of a variety of pointed observations. In contrast,
FIDEL is essentially a flux-limited catalog produced from a nearly
uniformly sampled map. We conclude that on the angular scale of the
BLAST beam sizes, one should anticipate that covariance with MUSIC
will provide a biased estimator of total intensity.

The FIDEL catalog, which has a substantially lower surface density,
does not show excess variance until scales of many arc minutes.  We
have tested whether this difference arises because FIDEL sources are
intrinsically different from MUSIC sources, or is instead simply a
feature of the shallower depth.  We repeat the tests shown in
Figure~\ref{fig:poisson_compare} for just the 1200 FIDEL sources that
lie in the MUSIC region and for several random subsamples of 1200
MUSIC catalog entries.  None of these curves is statistically
distinguishable from unity at any angular scale, and we conclude that
the Poisson variance associated with sampling MUSIC at the FIDEL
number density dominates over the correlations in galaxy locations
detected by MUSIC.

\section{Dissecting the Submillimeter background}
\label{sec:dissecting}

\subsection{Total Intensity}
\label{sec:total}

\begin{figure}[tp]
\centering
\includegraphics[width=\linewidth]{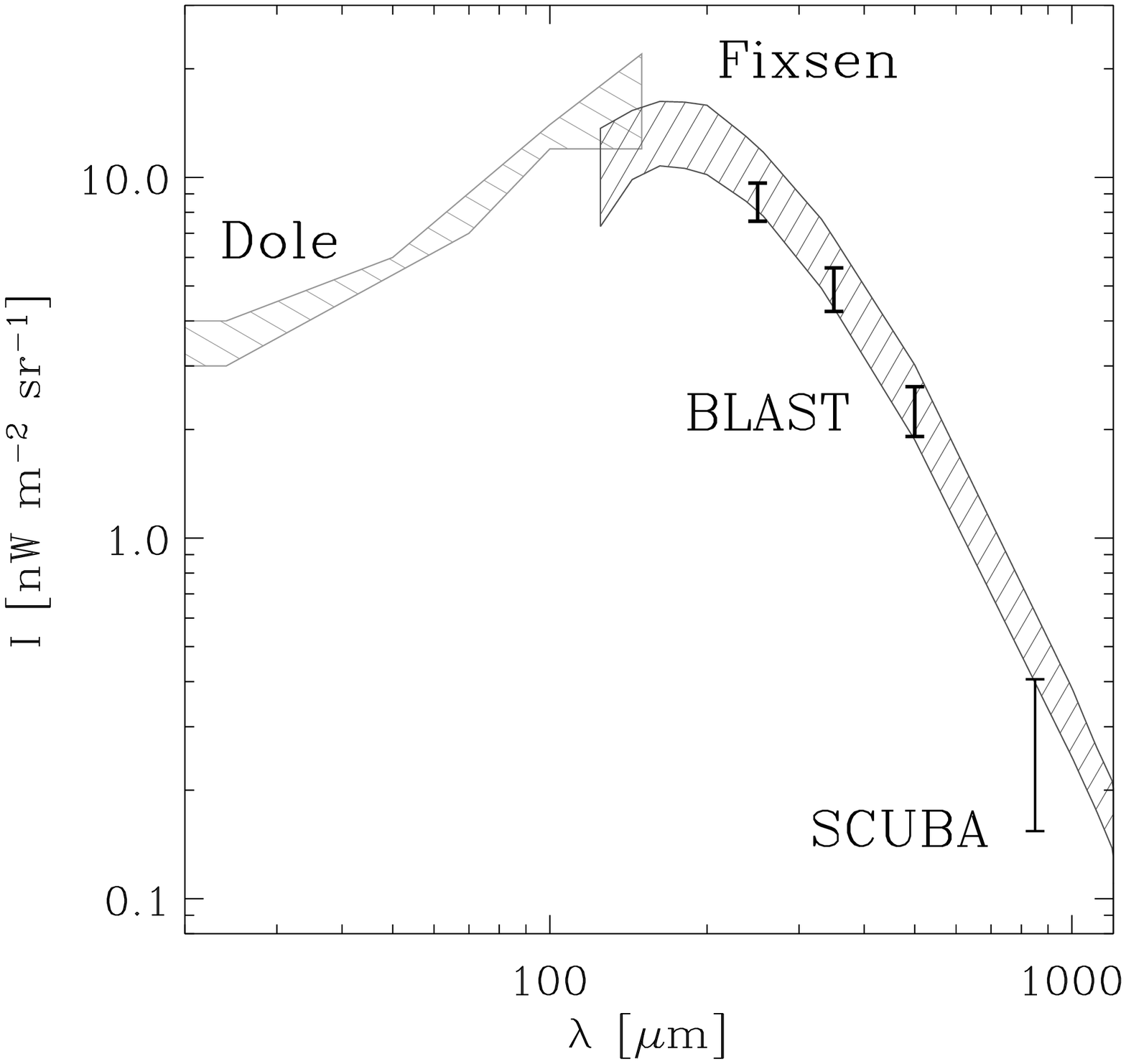}
\caption{ The total intensity of the far-IR background (CIB) in the
  \blast\ bands associated with 24\,\micron\ galaxies is plotted
  against wavelength. The \blast\ error bars include calibration
  uncertainties and are highly correlated between bands. The point at
  850\,\micron\ labeled SCUBA is a similar stack in GOODS-N using data
  described in \citet{borys2003} and \citet{pope2005}.  The hatched
  region marked ``Fixsen'' is the absolute spectrum of the CIB
  determined from FIRAS observations on {\sl COBE\/}
  \citep{fixsen1998}. The hatched region at $\lambda \le 160\,\micron$
  labeled ``Dole'' is the background reported by \citet{dole2006}
  using a similar stacking analysis of {\em Spitzer\/} data in
  GOODS-N.}
\label{fig:firb_plot}
\end{figure}

We stack the BLAST maps on the FIDEL 24\,\micron\ catalog, using the
methods described in \S\,\ref{sec:stacking}. Completeness-corrected
results are shown in Figure~\ref{fig:firb_plot} and listed in
Table~\ref{tbl:firb}, along with values of the CIB measured by
\citet{fixsen1998} and \citet{dole2006}. 
The BLAST measurements are consistent with having resolved 100\% of
the background, as measured by FIRAS. These numbers should be seen as
lower limits, though; see \S\,\ref{sec:redshift}. Even including
calibration uncertainty, the \blast\ total intensity values are twice
as precise as FIRAS. In the rest of this paper, total intensity of the
CIB refers to the \blast\ values listed in column~2 of
Table~\ref{tbl:firb}. We note that while the completeness of the FIDEL
catalog is uncertain (\S\,\ref{sec:stacking}), the correction to the
\blast\ background intensities is small ($<10\%$, see
\S\,\ref{sec:redshift}).

\begin{deluxetable*}{ccccc}
\tablewidth{0pt} \tablecaption{Corrected stacked
  intensities \label{tbl:firb}} \tablehead{ \colhead{} &
  \multicolumn{3}{c}{\bf \hrulefill\ \blast\ \hrulefill} & \colhead{}
  \\ \colhead{\bf Band} & \colhead{\bf Total} & \colhead{\bf Low-z} &
  \colhead{\bf High-z} & \colhead{\bf FIRAS} \\ \colhead{(\micron)} &
  \colhead{(\bgunits)} & \colhead{(\bgunits)} & \colhead{(\bgunits)} &
  \colhead{(\bgunits)} } \startdata 250 & $8.60 \pm 0.59\,(1.04)$ &
$5.18 \pm 0.45\,(0.69)$ & $3.42 \pm 0.37\,(0.50)$ & $10.4 \pm 2.3$
\\ 350 & $4.93 \pm 0.34\,(0.68)$ & $2.44 \pm 0.26\,(0.39)$ & $2.49 \pm
0.21\,(0.37)$ & ${\phantom 0}5.4 \pm 1.6$ \\ 500 & $2.27 \pm
0.20\,(0.36)$ & $0.89 \pm 0.15\,(0.20)$ & $1.38 \pm 0.13\,(0.22)$ &
${\phantom 0}2.4 \pm 0.6$ \\ \enddata \tablecomments{The quoted errors
  are measurement uncertainties only. The numbers in parentheses are
  errors including calibration uncertainties. The errors between the
  \blast\ bands are strongly correlated because \blast\ observes
  similar confusion-limited structure at all wavelengths. The Pearson
  correlation coefficients are $(\rho_{12},\rho_{13},\rho_{23}) =
  (0.70, 0.68, 0.70)$ for measurement-only uncertainties, where 1, 2,
  and 3 refer to 250, 350, and 500\,\micron, respectively. The
  coefficients for the full uncertainties are $(0.90,0.88,0.91)$,
  $(0.86,0.80,0.82)$, and $(0.84, 0.83,0.90)$ for the total, low-$z$,
  and high-$z$ stacks, respectively. We note that these numbers are
  slightly different from those in \citet{devlin2009}, due to a small
  update in the calculation methods.}
\end{deluxetable*}

\begin{figure}[tp]
\centering
\includegraphics[width=\linewidth]{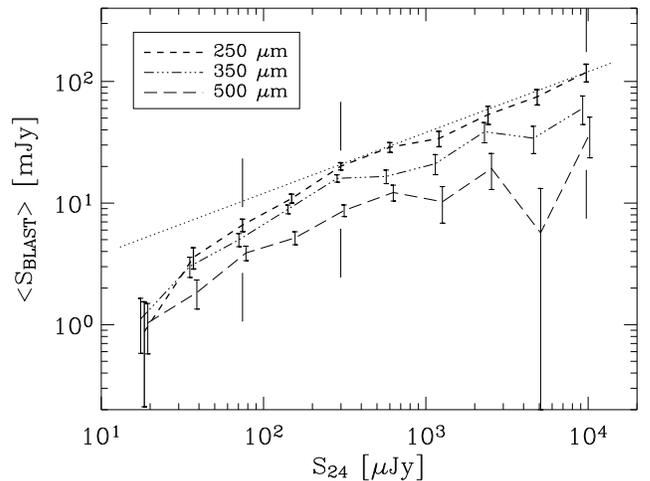}
\caption{The mean \blast\ flux density of FIDEL sources is plotted
  against 24\,\micron\ flux density for each of the three
  \blast\ bands.  The 350 and 500\,\micron\ curves have been displaced
  horizontally by $-5$ and $+5$\%, respectively, for visual clarity.
  The line near the uppermost curve is $S_\mathrm{\blast}\propto
  S_{24}^{1/2}$, shown for reference. The vertical lines indicate the MIPS
  flux densities for which mean spectra are plotted in
  Figure~\ref{fig:stack_spectrum}.  The number of FIDEL
  24\,\micron\ sources per bin decreases from ${\sim}\,2600$ at the
  faint end to 3 at the bright end. The lowest flux density bins are
  based on tentative 24\,\micron\ detections, and are therefore biased
  low (see \S\,\ref{sec:spurious}).}
\label{fig:aveflux}
\end{figure}

We examine the mean \blast\ flux density per source as a function of
24\,\micron\ flux density by dividing the FIDEL catalog into bins of
$S_{24}$ (Figure~\ref{fig:aveflux}).  The brightest sources in the
BLAST catalog are a factor of 10 times brighter than the mean
brightnesses of 24\,\micron-selected sources shown in this figure; it
is tempting to imagine that these curves extend to the right, but no
brighter 24\,\micron\ sources exist. We conclude that the
\blast\ sources have anomalous $S_\mathrm{\blast} / S_{24}$, yet the
full $S_\mathrm{\blast}\geq 3\,\sigma$ catalogs \citep{devlin2009} comprise only
10--15\% of the total intensity determined from stacking the full
FIDEL catalog, since the number density of these objects is low.

The data do not allow a {\em linear\/} relation between flux density
at 24\,$\mu$m and flux density in the \blast\ bands, such as any model
of the form 
\begin{equation}
S_\mathrm{\blast}=A\left(\frac{\lambda_\mathrm{\blast}}
  {24\,\micron}\right)^pS_{24}^\beta,
\end{equation}
with $\beta = 1$, would imply for any value of $p$.  This suggests
that the \blast-detected light comes from a distribution of sources
with different spectral energy distributions (SEDs), or with a range
of redshifts that varies significantly as a function of $S_{24}$.
This conclusion relies on the very large dynamic range, a factor of
500 in $S_{24}$, available from the FIDEL catalog. Even for a power
law relation, allowing $\beta$ to differ from 1, the data do not
support a constant value of $\beta$ except perhaps at high flux
densities. Figure~\ref{fig:aveflux} shows how the 250\,\micron\ curve
diverges from $\beta = 0.5$ at low flux densities.

\begin{figure}[tp]
\centering
\includegraphics[width=0.8\linewidth]{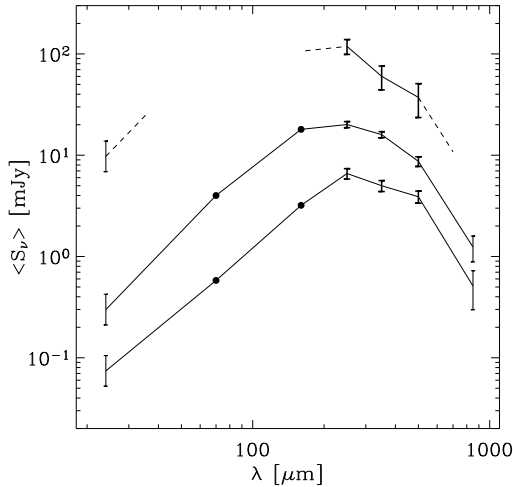}
\caption{Spectra of the mean flux density in \blast\ bands of sources
  of fixed 24\,\micron\ flux density.  The three curves correspond to
  the 24\,\micron\ flux densities indicated by the three vertical
  lines in Figure~\ref{fig:aveflux}. The point at 24\,\micron\ on each
  curve is the central flux density of the bin defining each
  sub-sample and the error bar indicates the width of that flux
  density bin.  The circles at 70 and 160\,\micron\ on the lower two
  curves are the mean flux densities in those bins determined by a
  similar stacking analysis in \citet[Figure~7]{dole2006}. 70, 160,
  and 850\,\micron\ flux densities are not available for the upper
  curve, so we plot dashed lines following the same slopes as the
  middle curve. A clear trend of increasing mean flux density and
  apparent temperature (spectrum peaking at shorter wavelengths) is
  observed in the \blast\ bands with increasing 24\,\micron\ flux
  density.}
\label{fig:stack_spectrum}
\end{figure}

Spectra of the mean flux density in \blast\ bands of sources at three
fixed 24\,\micron\ flux densities are shown in
Figure~\ref{fig:stack_spectrum}. 
This plot shows two features very
clearly: first, that the average BLAST flux density is positively
correlated with 24\,\micron\ flux density; and second, that fainter
24\,\micron\ sources appear cooler 
(as suggested by the shallower slopes through the
\blast\ wavelengths). This latter point is probably due to the fact
that higher-redshift sources have predominantly fainter
24\,\micron\ flux densities \citep[see, for example, the strongly
  non-Euclidean region of the source counts at $S_{24} < 1$\,mJy
  in][]{papovich2004} --- if the average rest-frame galaxy dust
temperature does not evolve appreciably, then the mean spectrum of
fainter/more distant objects will undergo greater redshift and hence
appear cooler. We note, however, that these curves are not well
described by a single-object SED, and are certainly averages over a
wide range of source types. The effects of a rough cut on redshift are
presented in the next section, and detailed \blast\ stacks as a
function of redshift are explored in \citet{pascale2009}.

\subsection{Division in Redshift}
\label{sec:redshift}

\begin{figure}[tp]
\centering
\includegraphics[width=0.8\linewidth]{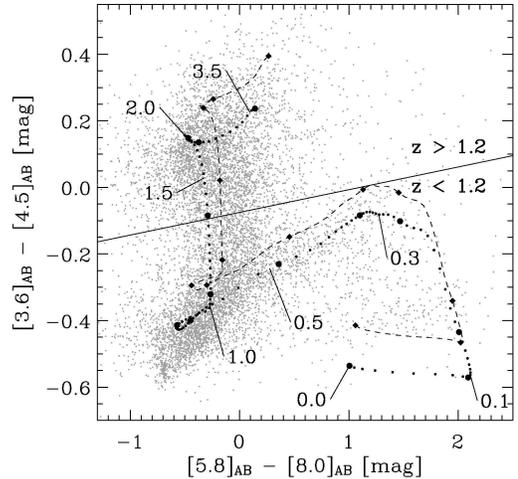}
\caption{IRAC color-color plot for sources in the FIDEL catalog (small
  grey dots). The sources lie in two partially-overlapping clouds. The
  black circles represent the colors of an SBc galaxy template
  \citep[VCC~1987]{devriendt1999} observed at a range of
  redshifts. The circles are linearly spaced in lookback time; the
  small circles are spaced by 0.1\,Gyr and the large circles by
  1\,Gyr. A range of redshifts are also indicated by the numbers along
  the track. 
  The dashed line is the track traced out by the star-bursting galaxy
  M82. The diamonds indicate intervals of 1\,Gyr, equivalent to the
  large circles for the SBc template.  The solid line indicates the
  color-cut used to separate high- and low-redshift sources ---
  sources above the line are mostly at $z > 1.2$ while sources below
  the line are mostly at $z < 1.2$. The choice of this line is
  described in \citet[supplement]{devlin2009}. We use
  H$_0=70.5$\,km\,s$^{-1}$\,Mpc$^{-1}$, $\Omega_\mathrm{M}=0.274$, and
  $\Omega_\Lambda=0.726$ \citep{hinshaw2009} to tabulate the lookback
  times.}
\label{fig:irac_cc}
\end{figure}

We use IRAC 3.6--8.0\,\micron\ flux densities of the FIDEL sources to
make a cut in a color-color plane to broadly classify the sources as
either high- or low-redshift (Figure~\ref{fig:irac_cc}).
The sources (small grey dots in the figure) mostly lie in two clouds,
one in the lower-left quadrant and the other in the upper-left
quadrant of the figure. For illustration, we over-plot the observed
colors of a galaxy template at a range of lookback times (\tlb),
ranging from 0--12\,Gyr (black circles). The template is the SBc
galaxy VCC~1987 from \citet{devriendt1999}, chosen because it is
publicly available and exhibits the major features of the color-color
space occupied by the FIDEL catalog.  We note that the clouds
correspond to regions of color-color space where this template galaxy
lingers for several Gyr. 
The track traced out by the local star-forming galaxy M82, thought to
be representative of high-redshift submillimeter galaxies
\citep{pope2008}, is also shown. Its IRAC colours as a function of
redshift are very similar to the SBc galaxy.

We adopt the line
\begin{eqnarray}
\lefteqn{\left([3.6] - [4.5]\right) =} \nonumber \\
&  & 0.0682 \times \left([5.8] - [8.0]\right) - 0.075,
\end{eqnarray}
where the quantities in square brackets are $AB$ magnitudes in the
IRAC bands, to divide the sources into two redshift bins. We have
checked this cut for the 4242 sources in the MUSIC catalog for which
spectroscopic or photometric redshifts are available, finding that
sources lie at redshifts higher and lower than $z=1.2$ above and below
the line, respectively.  The cut is remarkably sharp, with only 15\%
cross-contamination \citep[supplement]{devlin2009}. This color-color
cut is similar to the cut used by \citet[Figure~2]{yun2008} to
identify submillimeter galaxy counterparts.

\begin{figure}[tp]
\centering
\includegraphics[width=0.8\linewidth]{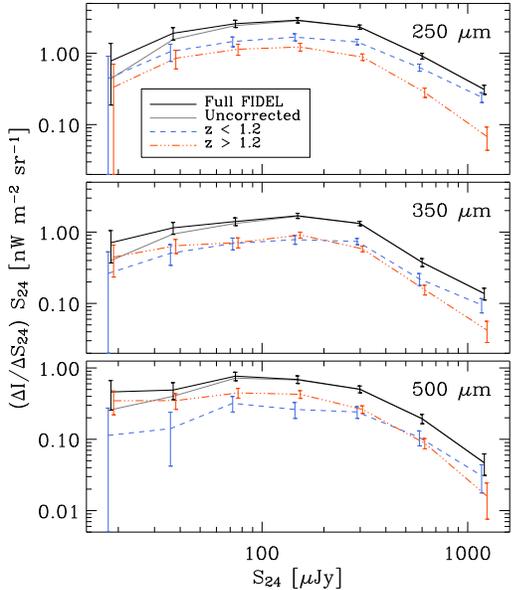}
\caption{ The completeness-corrected differential distribution of
  contribution to the CIB versus flux density of the corresponding
  24\,\micron\ galaxies. The result for the uncorrected differential
  stack is also shown for reference with a grey solid line to
  emphasize the fact that completeness only affects the two lowest
  bins.  The 24\,\micron\ sources are divided into high and low
  redshift bins, as described in \S\,\ref{sec:redshift}. The binning
  in 24\,\micron\ flux density is the same as in
  Figure~\ref{fig:aveflux}, with the 3 highest-flux density bins
  suppressed due to the small number of sources. The lowest
  24\,\micron\ flux density bin is comprised of tentative detections,
  and so is possibly less certain than the error bars indicate (see
  \S\,\ref{sec:spurious}). The low- and high-redshift curves have been
  displaced horizontally by $-3$ and $+3$\%, respectively, for visual
  clarity.}
\label{fig:diffstack}
\end{figure}

The differential contributions to the CIB from these ``high'' and
``low'' redshift sources as a function of 24\,\micron\ flux density
are shown in Figure~\ref{fig:diffstack}.  We also show the
corresponding curve for the total CIB (using the entire FIDEL catalog)
for reference.  Specifically, the stacks have been split into
logarithmic bins in flux density and the total stacked intensity is
determined for each bin, divided by the linear bin widths. We
therefore plot $(\Delta I/\Delta S_{24}) \, S_{24}$ as a function of
$S_{24}$, where $\Delta I$ is the total contribution to the CIB at a
given \blast\ frequency from the sources in each bin and $\Delta
S_{24}$ is the linear bin width.

The bulk of the CIB at 250 and 350\,\micron\ is clearly produced by
sources with $S_{24}\geq 60$\,\microjy, while the low-flux density
drop-off is less clear at 500\,\micron. This drop-off at low
24\,\micron\ flux density is due in large part to the number counts at
24\,\micron\ \citep[Figure~2]{papovich2004}, which are steepest at
${\sim}\,$200--1000\,\microjy\ and roll over significantly at lower
flux densities. There appears to be CIB intensity missing from our
measurements due to the flux density limit of the
24\,\micron\ catalog, in the sense that one would not expect strictly
zero \blast\ intensity in the next lower bin missing from each panel
in Figure~\ref{fig:diffstack}. We expect that this missing component
is small. The 24\,\micron\ number counts below
${\sim}\,20$\,\microjy\ are uncertain, but since the 2\,$\sigma$ upper
limits derived from FIRAS are not much above the values derived in
this analysis (Table~\ref{tbl:firb}), we conclude that there is no
hidden population of very faint 24\,\micron\ sources that are
important to the CIB at the \blast\ wavelengths.

Two trends in redshift are clear: first, that the fraction of the
total CIB due to high redshift galaxies increases with wavelength
through the \blast\ bands (compare the relative heights of the
dot-dashed and dashed curves in each panel); second, in all BLAST
bands, the relative contribution to the background produced by
high-redshift sources (i.e.\ observed to be colder) increases toward
fainter 24\,\micron\ flux densities (compare the heights of the
dot-dashed and dashed curves as a function of $S_{24}$).  This
analysis is consistent with our interpretation of
Figure~\ref{fig:stack_spectrum} described in \S\,\ref{sec:total}.

\begin{figure}[tp]
\centering
\includegraphics[width=0.9\linewidth]{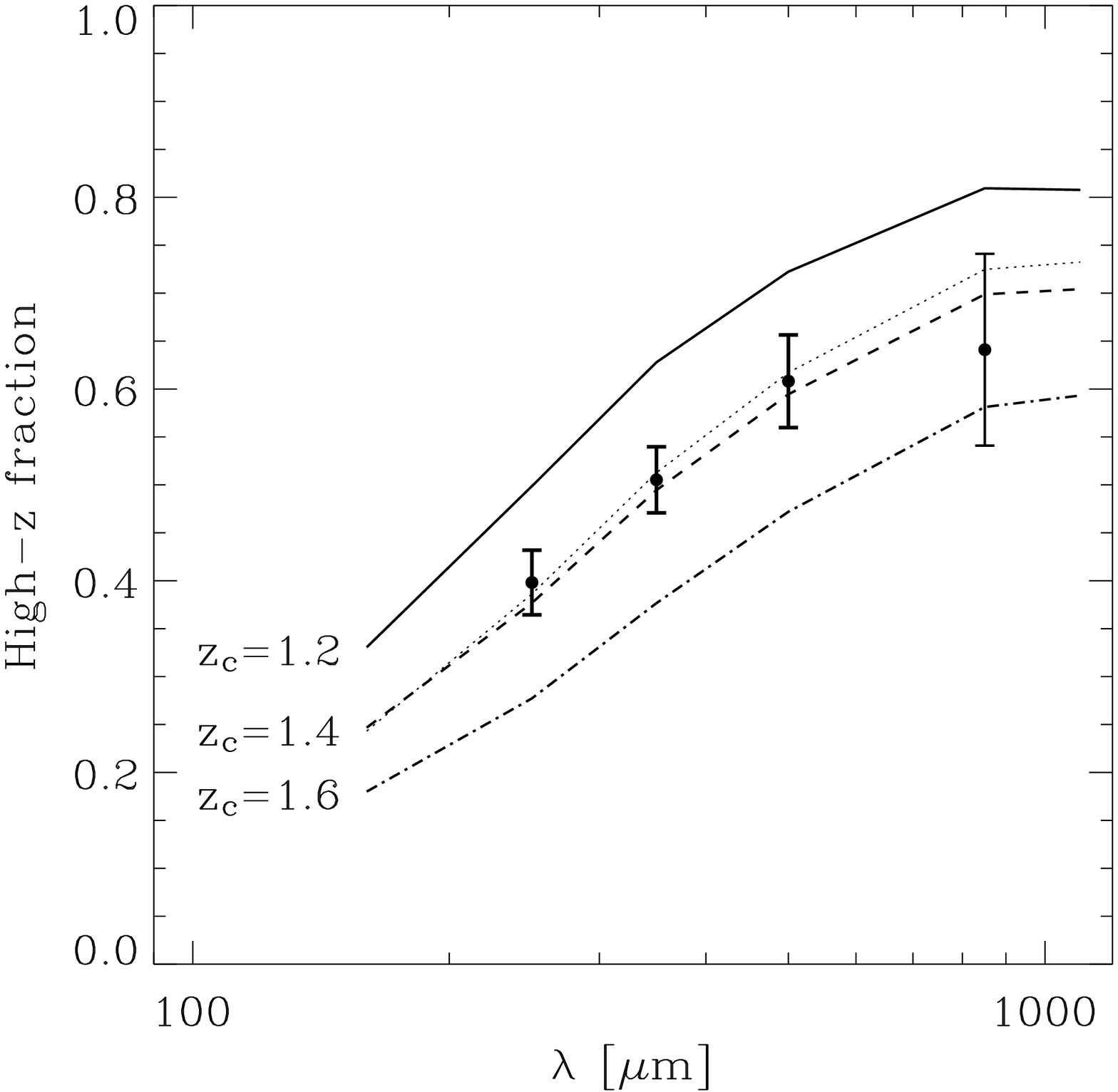}
\caption{The fraction of the 24\,\micron-driven CIB which comes from
  high redshift galaxies, as defined by the IRAC color-color cut shown
  in Figure~\ref{fig:irac_cc}, is plotted against wavelength for the
  three \blast\ bands and SCUBA. Over-plotted are predictions from the
  phenomenological evolutionary model of \citet{valiante2009}, for a
  range of redshift cuts. The solid, dashed, and dash-dotted curves
  are for a 24\,\micron\ catalog complete above
  $S_{24}=30$\,\microjy. The dotted curve shows how to the
  $z_\mathrm{c} = 1.4$ curve changes for a $S_{24} \geq
  10$\,\microjy\ catalog. $z_\mathrm{c}=1.2$ is clearly not a good fit
  to the data. We hypothesize that our estimate of $z_\mathrm{c}$
  corresponding to the redshift cut of \S\,\ref{sec:redshift} is
  biased low due selection effects. Note that the
  \blast\ uncertainties are highly correlated (Table~\ref{tbl:firb}).}
\label{fig:zratio}
\end{figure}

Figure~\ref{fig:zratio} shows the fraction of the CIB produced by the
high-redshift sample as a function of wavelength.  The curves are
predictions from the phenomenological evolutionary model of
\citet{valiante2009}.  The different curves show the ratios of total
intensities from galaxies brighter than $S_{24}=30$\,\microjy\ (solid,
dashed, and dot-dashed curves) and $S_{24}=10$\,\microjy\ (dotted
curve) from galaxies located at $z\geq z_\mathrm{c}$, as labeled. We
note that the curve for $z_\mathrm{c}=1.4$ is a substantially better
fit than the curve for $z_\mathrm{c}=1.2$, despite the fact that
photometric and spectroscopic redshifts in the MUSIC catalog suggest
that the IRAC color-color cut we have adopted corresponds to
$z_\mathrm{c}=1.2$.  Perhaps the subset of sources with redshifts from
this training set is biased, and the galaxies dominating the stacks
do, in fact, lie at slightly higher redshifts than the models
suggest. We note that a similar IRAC color-color cut proposed by
\citet{yun2008} is designed to identify counterparts to
850\,\micron-selected galaxies that lie at redshifts predominantly
$z>2$. Alternatively, there are parameters governing the density and
luminosity evolution of galaxies in the \citet{valiante2009} model
that could mimic this effect.

\subsection{Galaxies hosting an AGN}

We attempt to select for active galactic nuclei (AGN) in the FIDEL
sample based on color-color cuts. We attempted to use the power-law
selection suggested by \citet{donley2008}, but found that the majority
of sources, even those that look like a power law through the IRAC
bands, are rejected by the goodness-of-fit criterion, due to the small
relative errors in the IRAC flux densities. The majority of sources
that met both the power law index and goodness-of-fit criteria were
those with very poorly determined 5.8 and 8.0\,\micron\ flux
densities, and thus landed all over the color-color plane. Instead, we
make a cut on colors that encompasses all of the \citet{donley2008}
sources with minimal contamination from non-AGN:
\begin{eqnarray}
y & \ge & \phantom{-}1.21\, x -0.30 \\
y & \le & \phantom{-}1.21\, x +0.22 \\
y & \ge & -0.83\, x +0.16
\end{eqnarray}
where $y \equiv \log\left(S_{8.0}/S_{4.5}\right)$ and $x \equiv
\log\left(S_{5.8}/S_{3.6}\right)$. We find that 480 FIDEL sources meet
this criteria, which is similar to the expected number of
X-ray--detected AGN, based on the source density reported by
\citep{luo2008}. The average \blast\ flux density at the positions of
the IRAC color-selected sources is $7.5\pm1.1$, $7.4\pm0.9$, and
$6.0\pm0.8$\,mJy at 250, 350, and 500\,\micron, respectively. These values
are significantly larger than for the full FIDEL list,
$4.6\pm0.2$, $3.6\pm0.2$, and $2.3\pm0.2$\,mJy.

\begin{figure}[tp]
\centering
\includegraphics[width=0.9\linewidth]{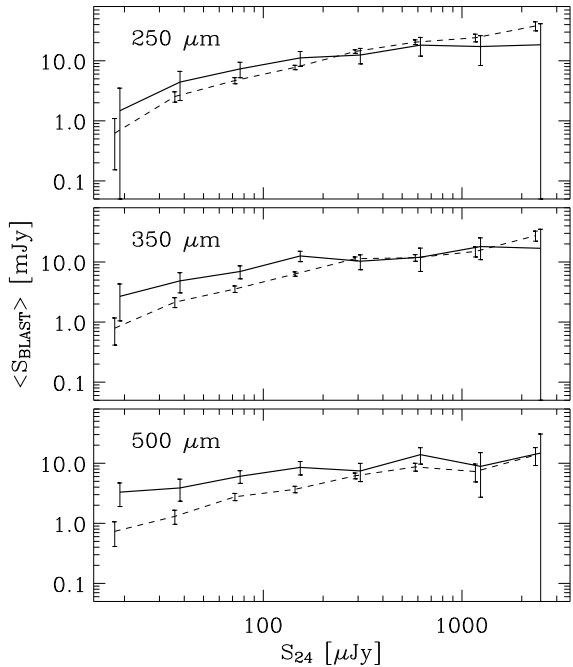}
\caption{The average \blast\ flux density of 24\,\micron\ sources
  classified as AGN (solid lines) and for the full FIDEL list (dashed
  line). The classification is a color-color cut based on the power
  law seletion of \citet{donley2008}. The lowest flux density bins are
  based on tentative 24\,\micron\ detections, and are therefore biased
  low (see \S\,\ref{sec:spurious}). The curves are shifted
  horizontally by $+5$ and $-5$\% for clarity.}
\label{fig:fidel_agnstack}
\end{figure}

Figure~\ref{fig:fidel_agnstack} shows the average \blast\ flux density
binned by 24\,\micron\ flux density for the FIDEL sources labeled as
AGN (solid line) and for the full set (dashed line). We see that
across all \blast\ bands, the AGN are brighter than average at all but
the greatest 24\,\micron\ flux densities, with the ratio (averaged
over all 24\,\micron\ flux densities) increasing from 1.6 to 2.6, from
250 to 500\,\micron.  The fact that this fraction increases at longer
wavelengths (under the assumption that higher-redshift sources should
appear cooler) is puzzling since the space densities of both AGN and
the ultra-luminous star-forming galaxies selected at
850\,\micron\ grow quickly to redshifts $z\sim2$--2.5
\citep[e.g.][]{chapman2005,wall2005,wall2008}. Either the dust in
star-forming galaxies is generally warmer than in the hosts of AGN, or
the total background produced by purely star-forming galaxies occurs
at lower redshifts. This latter explanation is more likely, since it
is now believed that the bulk of the total stellar mass in the
Universe was formed in more numerous, less luminous galaxies at lower
redshifts than those detected in longer-wavelength submillimeter
surveys \citep[e.g.][]{perez2008}.

For comparison, we also stack on catalogs of AGN selected by the {\sl
  Chandra}\/ X-ray Observatory and by optical
variability. The X-ray sample of \citet{luo2008} contains 462 sources,
164 of which (ignoring upper limits) fall in the region of the
$R$-band vs. soft X-ray flux plot that is thought to select for AGN
\citep[Figure~12]{luo2008}. The average \blast\ flux density at the
positions of the X-ray--selected AGN is $7.3\pm1.9$, $7.5\pm1.5$, and
$5.3\pm1.3$\,mJy at 250, 350, and 500\,\micron, respectively, similar to
the values listed above. We note that the stack on the full list of
462 is marginally larger at 250\,\micron\ ($9.4\pm1.2$\,mJy), although
nearly identical at 350 and 500\,\micron\ ($7.6\pm1.9$ and
$5.1\pm0.8$\,mJy). The optical variability sample of
\citet{trevese2008} contains 132 objects, with average flux densities
of $9.1\pm2.1$, $7.1\pm1.7$, and $4.4\pm1.4$\,mJy at 250, 350, and
500\,\micron, respectively. 

Our measurements are tantalizing since it is one of the major goals in
modern cosmology to establish an evolutionary connection between the
formation of the most massive galaxies and the growth of black
holes. However, we warn the reader that the color-color selection
described above may be significantly contaminated by star-forming
galaxies, which would introduce strong biases such that the observed
trends here may be unrelated to the underlying evolution in the AGN
and star-forming galaxy populations. A catalog of AGN which is
complete and free of interlopers would allow us to improve our
understanding of the relative importance of AGN in the formation of
the CIB.

\subsection{{\em BzK\/} Galaxies}

\citet{daddi2004} have shown that selecting galaxies that are dimmer
in $z$-band than they are in $B$ and $K_\mathrm{s}$ yields a list which
contains many star forming galaxies.  Specifically, \bzk\ galaxies are
those for which
\begin{equation}
\Delta_{BzK}\equiv (z-K)_{AB} - (B-z)_{AB} > -0.2,
\end{equation}
which tends to select star forming galaxies at $z \geq 1.4$.  Their
association with ULIRGs has been studied \citep{daddi2005}, and
\citet{hartley2008} conclude from their mild clustering properties
that they are associated with dark matter halos.

We use photometry from the 1548 FIDEL sources with MUSIC catalog
identifications to calculate $\Delta_{BzK}$.  There are 388
\bzk\ galaxies in the region, while the BLAST $5\,\sigma$ catalog
contains only a handful of entries, so we know that the
\bzk\ criterion is not selecting BLAST sources in general.  Separating
all 1548 FIDEL galaxies into bins by $\Delta_{BzK}$ and stacking on
the galaxies in each bin we find only a very weak dependence of mean
BLAST flux density on $\Delta_{BzK}$.  \bzk\ galaxies, i.e.\ sources
with $\Delta_{BzK} > -0.2$, are mildly brighter than others; the
FIDEL/MUSIC \bzk\ sources have average BLAST flux densities 1.3--1.7
times greater than for the full list, and produce $32\pm6$, $34\pm7$,
and $42\pm11$\% of the total BLAST intensity at 250, 350, and
500\,\micron, respectively. It is clear that \bzk\ samples make an
important contribution to the CIB, but this is largely because they
are so numerous.

For comparison, \cite{takagi2007} find that \bzk\ galaxies produce 9\%
of the CIB at 850\,\micron, as measured by FIRAS\@. If we restrict our
sample to galaxies with $K_S \le 22.9$, the limiting magnitude of the
\citeauthor{takagi2007} sample, we find that the \bzk\ galaxies make
up 12, 10, and 12\% of the CIB at 250, 350, and 500\,\micron,
respectively, comparable to \citeauthor{takagi2007}

\subsection{Spurious Detections in FIDEL Catalog}
\label{sec:spurious}

As described in \S\,\ref{sec:fidel}, detections in the FIDEL catalog
below 30\,\microjy\ are considered tentative. This affects the lowest
flux density bin in Figures~\ref{fig:aveflux}, \ref{fig:diffstack},
and \ref{fig:fidel_agnstack}. Since Figures~\ref{fig:aveflux} and
\ref{fig:fidel_agnstack} are average flux densities, the effect of
spurious sources (which are presumably uncorrelated with the
\blast\ maps) is to lower the estimate, so the lowest flux density bin
is probably biased low. The uncorrected curves in
Figure~\ref{fig:diffstack} are {\em sums}\/ over the FIDEL sources, so
spurious sources do not affect the measurement, and there is no
bias. Spurious sources would affect the completeness correction,
however, and thus the curves labelled ``Full FIDEL'' should also be
regarded as biased low. In addition, since the quoted uncertainties
take into account the number of sources included in the stack, the
error bars are likely smaller than they should be.

\section{Conclusions}

For the first time, we trace the full submillimeter intensity of the
far-IR background using a specific catalog of 24\,\micron-selected
galaxies. We use the properties of these 24\,\micron\ galaxies to
analyze the composition of the CIB. We have determined that the
average submillimeter flux density varies non-linearly with
24\,\micron\ flux density; these results require that the underlying
averaged SED evolves with redshift. We make precise measurements of
how the fraction of the CIB produced at higher redshifts varies with
observed wavelength through the submillimeter spectrum. These values
are lower than models predict. We determine the total submillimeter
intensity from unresolved AGN using an IRAC color-color selection,
finding weak evidence that they produce a non-negligible fraction of
the CIB, and a relatively larger fraction in the longer-wavelength
BLAST bands. This result could be demonstrating that the epoch of peak
star-formation is at lower redshifts than the peak in AGN space
density, which might be expected if most stars form in
lower-luminosity, lower-redshift galaxies than the ultra-luminous
star-forming galaxies selected at 850\,\micron. However, this result
depends strongly on the AGN selection criterion which may contain many
sources without an AGN, and which is also expected to be incomplete at
higher redshifts. We also stacked at the positions of \bzk\ galaxies,
finding that they contribute 32--42\% of the CIB in the BLAST bands,
consistent with the contribution found by \citet{takagi2007} at
850\,\micron\ when the depths of the optical catalogs are considered.

\acknowledgments

Thanks to Alex Pope for providing the SCUBA GOODS-N map, and to
J.~E.~G.~Devriendt for providing model galaxy spectra. We acknowledge
the support of NASA through grant numbers NAG5-12785, NAG5-13301, and
NNGO-6GI11G, the NSF Office of Polar Programs, the Canadian Space
Agency, the Natural Sciences and Engineering Research Council (NSERC)
of Canada, and the UK Science and Technology Facilities Council
(STFC). This research has been enabled by the use of WestGrid
computing resources. This work is based in part on observations made
with the Spitzer Space Telescope, which is operated by the Jet
Propulsion Laboratory, California Institute of Technology under a
contract with NASA.

\bibliographystyle{apj} \bibliography{apj-jour,refs}

\end{document}